\documentclass[a4paper]{article}

\usepackage{INTERSPEECH2022}
\usepackage{algpseudocode}

\title{Sotto Voce: Federated Speech Recognition with Differential Privacy Guarantees}
\name{Michael Shoemate$^1$, Kevin Jett$^2$, Ethan Cowan$^1$ 
\\ Sean Colbath$^2$, James Honaker$^1$, Prasanna Muthukumar$^2$ }
\address{
  $^1$Harvard University, Cambridge, MA, USA\\
  $^2$Raytheon BBN Technologies, Cambridge, MA, USA}
\email{shoematem@seas.harvard.edu, kevin.jett@raytheon.com,ecowan@g.harvard.edu, sean.colbath@raytheon.com, honaker@seas.harvard.edu, prasanna.muthukumar@raytheon.com }

\begin{document}

\maketitle
\begin{abstract}
Speech data is expensive to collect, and incredibly sensitive to its sources. It is often the case that organizations independently collect small datasets for their own use, but often these are not performant for the demands of machine learning. Organizations could pool these datasets together and jointly build a strong ASR system; sharing data in the clear, however, comes with tremendous risk, in terms of intellectual property loss as well as loss of privacy of the individuals who exist in the dataset. In this paper, we offer a potential solution for learning an ML model across multiple organizations where we can provide mathematical guarantees limiting privacy loss. We use a Federated Learning approach built on a strong foundation of Differential Privacy techniques. We apply these to a senone classification prototype and demonstrate that the model improves with the addition of private data while still respecting privacy. 
\end{abstract}
\noindent\textbf{Index Terms}: speech recognition, differential privacy, acoustic model

\section{Introduction}
Imagine this scenario: Organization X wants to build a speech recognizer for Dari, a low-resourced language. 
X does not have enough Dari data to train a high-accuracy recognizer. 
X knows that organizations Y and Z have also collected Dari datasets, but neither of these datasets are large enough to build a high-accuracy recognizer. 
All three organizations can benefit from building ML models on the pooled data, however, legal and ethical risk impedes directly sharing this data.

Once data is released, it can never be taken back. 
Organizations X, Y, Z will therefore be giving each other access to their data for perpetuity.
One could argue that licenses and legal protections could limit how long each actor can cache the others' data. 
Unfortunately, this protection is only as secure as the least compliant actor; it relies purely on the willingness of the various organizations to follow the law. 
Even an organization that is ethically sound today may experience acquisitions and mergers that change their constitution.
Even sharing aggregate statistics or models about the data comes with privacy concerns.  
Federated Learning\cite{kairouz2021advances} is a widely-used approach for learning models on data from multiple sources. 
Oftentimes, federated learning schemes consist of a central parameter server that aggregates neural network gradients from multiple sources.
Each organization contributes neural network gradients (never the data itself) and receives updates for their own local model.
Unfortunately, a malicious actor can still extract sensitive information from the gradients.   
The two most likely attacks involve membership inference and model inversion. 
An attacker can use membership inference to detect the presence of an individual in another organization's data. 
Model inversion is a more serious attack where the trained model is \emph{inverted} to reconstruct the dataset it was trained on.   Even gradients have been shown to be susceptible to attacks that reconstruct the underlying data.
With speech data, the risk of losing your privacy due to attacks is particularly severe. 
To start, individuals can lose sole command of their voice, since speech synthesizers can be built with as little as 30 minutes of speech data\cite{black2015random}.
Unlike leaked passwords, there is no chance of resetting one's own voice and starting afresh. 
More importantly, compared to text mediums, speech is considered an impermanent medium, so audio recordings have a greater tendency to violate our privacy norms.
High-quality speech models need to train on large amounts of speech data, so model inversion attacks could expose sensitive speech. 

One approach that is often proposed for releasing private data is to transform the dataset into a set of features and then release the features instead. 
This solution is na\"ive.
In the federated context, featurization does not lend any protection against membership inference attacks, since the attacker --- being part of the federation scheme --- has access to the featurizer.
Also, most forms of feature transformations do not have mathematical guarantees that they are one-way functions. 
Without such guarantees, it is not possible to ensure that the featurization cannot be reversed, i.e. that the dataset cannot be reconstructed from the features.
Generally speaking, a good featurizer preserves information in the input dataset to facilitate learning, which makes it, by its very nature, susceptible to inversion.
We need strong guarantees that private data will never be leaked.  
While the obvious solution might be to refuse to share any data whatsoever, a more interesting solution would be to find a way to share data while still bounding the privacy loss.

In this paper, we describe a Federated Learning technique for building an acoustic model built on the foundations of Differential Privacy (DP)\cite{dwork2014algorithmic}. DP provides us mathematical guarantees limiting privacy loss, and also gives us adjustable parameters that allow us to trade off privacy for performance. The end result is an acoustic model that is able to learn from multiple sources while still respecting the privacy of each source's data. 

\begin{figure}[htbp]
  \centering
  \includegraphics[width=0.95\linewidth]{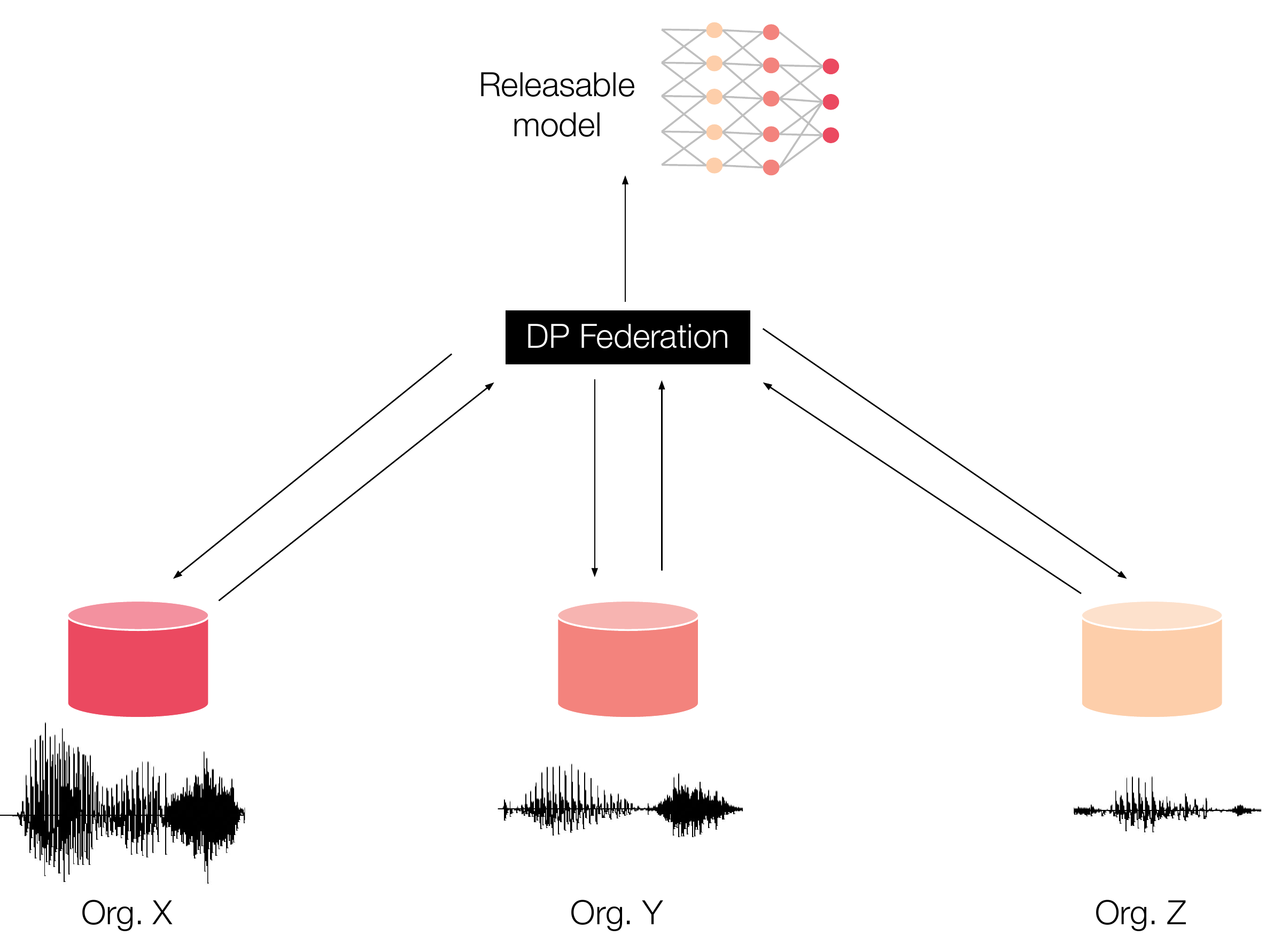}
  \vspace{-3mm}
  \caption{DP-based Federated learning techniques allow us to build ML models across organizations while still preserving privacy to the extent guaranteed by DP}
  \label{fig:federated_dp}
  \vspace{-7mm}
\end{figure}

\section{Related Work}
Speech processing and privacy-preserving computation have each independently had a long and storied history, but there has been little work at the intersection of the two. The earliest of these is the work by Smaragdis and Shashanka\cite{smaragdis2007framework}. The authors assume two parties, one of whom has a private dataset and another who has an HMM-based trained speech recognition system. They propose a technique based on Secure Multi-party Computation (SMC) where the two parties can perform a speech recognition task without having to share the dataset openly or reveal details about the HMM recognizer. An alternate approach to the same problem is described in Pathak et al\cite{pathak2011privacy} which uses Homomorphic Encryption instead of SMC. The major difference between our work and these two papers is that these papers (and other along the same vein\cite{pathak2011privacya,pathak2013privacy,ahmed2020preech}) deal with privacy preservation at \emph{inference} time, and they do not result in private information shared between parties. In contrast, our work targets model \emph{training}, and the purpose is to build models that may be shared with interested parties with limited risk to individuals used to train the model. 

In more recent years, with the advent of the European Union's General Data Protection Regulation, there has been more attention focused on protecting privacy in training in addition to privacy in inference. A prime example of this is the Voice Privacy Initiative and Challenge\cite{tomashenko2020introducing,tomashenko2022voiceprivacy}. The focus of these has been voice \emph{anonymization} i.e. removing aspects of identity from speech before putting it through the speech processing system. While this is a worthwhile task, our approach is complementary while still trying to reach the same goals. 

Our approach relies primarily on federation and differential privacy. Federated learning approaches to speech recognition are relatively popular and common, but the main criticism lobbed at such approaches to privacy is that the resulting model can be subjected to a membership inference or model inversion attack. These attacks could result in revealing private information about the speakers used in training. However, combining federation with differential privacy effectively blocks such attacks. Therein lies the novelty of our work.

\section{Differential Privacy}

A detailed, easy-to-read, non-mathematical intro to Differential Privacy is available here: \cite{wood2018differential}. However, in the interest of keeping this paper self-contained for the speech researcher, we briefly describe DP here.


To get a sense for Differential Privacy, imagine a tabular dataset where each row contains information about a specific individual. We would like to keep this dataset private, i.e. the information pertaining to specific individuals should not be made public. The core idea of DP is that anyone is allowed to get \emph{global} statistics such as means and variances about the dataset, but it is not possible to get information about a \emph{specific} individual. For instance, if one column of our dataset lists incomes, it should be possible for an interested person to find out the average income of everyone in the dataset. But it should be impossible to identify the income of a single person. 

To understand DP, it is important to first understand \emph{adjacent datasets}. An adjacent dataset is any dataset that differs from your dataset by a single individual. In our tabular dataset example, let's assume that our dataset $D$ has a row that contains information about a person, Jean. Then one adjacent dataset $D'$ to $D$ would contain every row in $D$ except the row with Jean's information. You can construct other adjacent datasets to $D$ by dropping other rows instead of the one with Jean, adding a new individual, or even by editing an individual.

The core principle of DP is that a DP statistical estimate (like a mean) computed on your dataset should be indistinguishable from the same estimate on any adjacent dataset.
Statistical estimates that follow this principle are \emph{mathematically guaranteed} to reveal at most a bounded amount of information about any specific individual in the dataset. 
To define this mathematically, first let $M(\cdot)$ be a randomized statistical estimator, like a DP mean. 
For $M(D)$ to be differentially private, then for any set of possible outputs $S$, the probability of the statistic on your data being in $S$ must be indistinguishable from the probability of the statistic on any adjacent dataset being in $S$. ``Indistinguishability" is quantified by $\epsilon$ and $\delta$.
Or equivalently:
\begin{equation}
Pr[M(D) \in S] \leq exp(\epsilon)\cdot Pr[M(D') \in S] + \delta
\end{equation}
Specifically, the distributions of $M(D)$ and $M(D')$ may differ by at most a multiplicative factor $\epsilon$ and an additive factor $\delta$ for $M(D)$ to qualify as $(\epsilon, \delta)$-differentially private.
This definition gives us a very robust and practical way to quantify privacy, but it also raises a problem for statistics we are familiar with, like the mean, that are deterministic. 
DP methods substitute traditional statistical estimators with randomized estimators calibrated to add the smallest amount of noise necessary to satisfy the definition of privacy.
The challenge of DP is to find unbiased estimators for statistics that analysts need that introduce the least variance possible, and yet can be calibrated to satisfy any given $(\epsilon, \delta)$.


To construct a function that obeys the principles of DP, we begin by analyzing the \emph{Global Sensitivity} (GS) of the function that we are interested in computing. 
Global sensitivity is an upper bound on how much any one individual can cause the function to change. 
The global sensitivity of many queries, like mean income, is infinite. 
There is no upper limit for an individual's income! 
To bound the sensitivity we must introduce some bias by clamping, replacing any value outside of $[L, U]$ with the nearest bound.
The sensitivity of this query on a clamped dataset known to have $N$ records is then $(U - L)/N$.

Now that we are familiar with global sensitivity, we put the pieces together to make a DP mean estimator. 
The Laplace distribution, scaled by $GS/\epsilon$, is tailored to and satisfies the definition of privacy. 
Intuitively, higher sensitivity queries result in more noise.
When the privacy guarantee is strengthened, epsilon gets smaller, and noise scale increases. 
This forms a simple $(\epsilon, 0)$-DP mean estimator:
\begin{equation}
 DP\ mean = \frac{\sum^N_i clamp(x_i, L, U)}{N} + Lap(\frac{U - L}{N \epsilon} )
\end{equation}

Another common noise distribution is the Gaussian, which admits looser $(\epsilon, \delta)$ guarantees.
An essential property of differential privacy is composition. 
When multiple DP releases are made, their privacy losses add up, or compose, linearly.
Every DP release is accounted for; even if we were to release multiple randomized estimates for the same value.
A high-level perspective on differential privacy is that it relates close datasets (by edit distance) to close queries (by sensitivity) to close distributions (by $\epsilon$, $\delta$)\cite{gaboardi2020programming}.
This allows us to relate real-world datasets with known limits on user contribution to rigorous privacy claims in terms of $\epsilon$ and $\delta$.

Both $\epsilon$ and $\delta$ give a privacy versus utility trade-off. 
There is no single parameter setting that is right for every application, and the privacy parameters should be tuned relative to the level of exposure acceptable to individuals in the dataset. 
The \emph{privacy budget} is the greatest acceptable choice of parameters.



\section{Differentially Private SGD}

The gradient of the loss function with respect to each of the parameters is a convoluted transformation from the input dataset to another dataset of gradients.
Each row of this hypothetical dataset of gradients contains an individual's private gradient updates to all network parameters.
Notice that adjusting one individual on the original dataset may only influence the same individual in the dataset of gradients. 
This means the gradient function is ``stable", and we can reuse our DP mean estimator to find a private gradient update, with some slight adjustments--
It is more efficient to clamp the $L_2$ norm of each individual's vector of gradients than it is to clamp each parameter's gradient individually.
We also switch to gaussian noise, because there are tighter approaches for composition over a large number of queries.
We also exercise DP's immunity to postprocessing. If a data release is $(\epsilon, \delta)$-DP, then any pure function evaluated on the release remains $(\epsilon, \delta)$-DP.
Rescaling the gradient by the learning rate and updating the weight matrix are postprocessing.
Since each step is DP, the entire SGD algorithm is also DP by composition\cite{abadi2016deep}.

In summary, we add these steps to standard SGD:

\begin{itemize}
\item \textbf{Clamp} gradients that exceed the $L_2$ norm bound
\item \textbf{Add noise} via the Gaussian mechanism
\item \textbf{Account} for the amount of privacy budget we spend
\end{itemize}


\section{Federated Learning}
Our goal in this work is to architect a scheme where multiple workers can jointly build an ASR model with privacy protections afforded to individuals in the training data, and without pooling the data openly. 
We realized this scheme with the \texttt{DistributedDataParallel} (DDP) module in PyTorch\cite{paszke2019pytorch}.
This module supports training over multiple workers on isolated hardware instances, each having access to a different private sequestered dataset, via TCP initialization.
Each worker maintains their own model replica, and conducts SGD training in tandem with the other workers.
In each worker, as the backpropagation algorithm runs, DDP eagerly synchronizes the gradients with a central worker, averages the gradients, and pushes the averaged gradients back to each worker.
Since all workers are initialized with the same weights, and take steps with the same federated gradients, they remain synchronized.
This lends significant flexibility, as each worker can individually choose privacy budget allocations and parameter settings.
It is even possible for parties to contribute non-noisy gradients during this training process, if they have datasets that do not have privacy concerns.

We wrote software that replaces the gradients with differentially private gradients before DDP communicates with other workers.
This was accomplished by registering hooks that fire when a gradient is backpropagated to any parameter tensor.
When this hook fires, gradients for each training instance are reconstructed from cached activations and backpropagations.
The hook returns a DP mean computed on these instance-level gradients.
Since the gradient privatization hook on tensors is invoked before the DDP hook shares gradients, DDP shares a DP gradient release with the other workers.
The federation procedure is solely a function of DP releases, and not underlying sensitive data, so the entire federation procedure is postprocessing. 
We use the openDP library\cite{gaboardi2020programming} from the OpenDP project to implement our algorithms\cite{opendpurl}. All our code has been integrated into this library and is already available for public use.   



\section{Senone Classification}

Implementing DP-SGD and its variants involves changing low-level aspects of NN training such as modifying the way gradients are computed during backpropagation. As a result, we have created our own implementations of many fundamental operations involved in SGD. While these are written to be fast and efficient, they have not benefited from the years of optimization that standard routines in PyTorch have enjoyed. As a result, NN training is slower with DP than without. This is to be expected when working on the frontier of research. DP training also requires direct computation of gradients for each individual, which incurs significant added overhead for all DPSGD implementations. 

Slowing down training also has the effect of slowing down research progress. To mitigate this effect, we run all our experiments on a \emph{senone classifier} rather than a full acoustic model. Senones are clustered triphones\cite{hwang1992subphonetic}, and senone classification is a slightly simpler task than full acoustic modeling. It is therefore a useful proxy for acoustic modeling while being faster and enabling us to run more experiments. 

The senone classifier we will be using for all our experiments is a neural network with a single hidden layer of LSTMs. The inputs are 13-dimensional Mel Frequency Cepstral Coefficients while the LSTM hidden layer has 200 units. The number of output classes a.k.a number of senones is 9096. The resulting NN therefore has a 13x200x9096 architecture. This classifier is obviously far from being state-of-the-art. However, for the work in this paper, we needed a small, easy-to-run prototype, and for that purpose, this network served us well. 

The primary dataset we will be using is Librispeech\cite{panayotov2015librispeech}. While the full Librispeech corpus is 1000 hours, we will be using smaller subsets of the corpus, a little over an hour each. 
The reason for the smaller subsets is to create a  more realistic situation for our cooperative federated learning use case. 
That being said, the DP guarantees would have improved substantially with the full dataset. The expected improvement is approximately linear relative to an increase in the number of unique individuals in the dataset.

\begin{table}[htb]
    \centering
    \begin{tabular}{c c c c}
    \toprule
    Data Set & Source & Duration & No. of \\
     & & (hrs:min) & speakers\\
    \midrule
    Public Data     &  Librispeech & 1:12 & 6\\
    Private set 1 & Librispeech & 1:26 & 11 \\
    Private set 2 & Librispeech & 1:25 & 8 \\
    Private set 3 & Speaker JB & 0:30 & 1\\
    \bottomrule
    \end{tabular}
    \caption{Details of the datasets used in our experiments}
    \label{tab:configuration}
    \vspace{-6mm}
\end{table}

As seen in Table~\ref{tab:configuration}, we created a `public' subset of Librispeech and two `private' subsets. In addition to these, we also created a new private dataset of speaker JB\footnote{JB is a well-known expert in the DP community, and this corpus was created by transcribing one of their lectures}. The Librispeech corpus is very high quality, studio-recorded speech spoken by professional voice talents. DP works on the principle of only revealing aspects of the data that are common across individuals. We therefore wanted to construct a private dataset that was pathologically different to test DP's abilities in privatization. Unlike Librispeech, JB's corpus is spontaneous speech with disfluencies recorded in a classroom environment, and JB is not a professional voice talent. JB is therefore such a strong outlier that their data will be more susceptible to attacks. 

\section{Results}
We trained on the public dataset for 51 epochs using conventional training schemes to generate a warm start. We then trained the model through Federated DP-SGD over the three private datasets for 14 epochs. We experimentally trained the senone classifier with a step $\epsilon$ of 100, $\delta$ of 1e-6, learning rate of 1e-4, and clipping bound of 1. Gaussian noise works out to a SD of .098.
In the DP community, the total composed privacy parameters are considered prohibitively large for model release, and we likewise do not recommend privacy parameters of this magnitude. 
Nevertheless, it is interesting to see that even when formal guarantees are significantly weakened, a practical membership attack is still thwarted.
Due to these large privacy parameters, we still consider our work to be in early stages.
These privacy parameters can be significantly improved using techniques in the discussion section.


Accuracy results from our experiments are shown in Table~\ref{tab:scores}. In addition to the training sets we described in the previous section, we also held out parts of these sets as test sets. One of these is from Librispeech while the other is a hold out from the JB set. The first row of the table shows the performance of a model trained purely on the public dataset.

\begin{table}[htb]
    \centering
    \begin{tabular}{c l c}
    \toprule
    Training Set & Test Set & Accuracy \\
    \midrule
    Public Data     &  Librispeech & 38.0 \\
      &  JB dataset & 23.1 \\
    \midrule
    Public + open private data & Librispeech & 47.1 \\
     & JB dataset & 46.0 \\
    \midrule
    \textbf{Public + DP private data} & \textbf{Librispeech} & \textbf{38.6} \\
      & \textbf{JB dataset} & \textbf{22.0} \\
    
    \bottomrule
    \end{tabular}
    \caption{Senone Classification Accuracies}
    \label{tab:scores}
    \vspace{-8mm}
\end{table}

The second row of the table shows the situation where the model was trained pretending that the private data was in fact public, i.e. all the data was lumped together and trained like a standard ASR system would be trained. As expected, there is a large jump in performance, especially on the JB test set. If we were to view this result through the narrow lens of ASR performance, we might think that this is a great result. From a privacy preservation point of view though, the result is worrisome. If we were to release this trained model to the public, it would be very easy to identify that JB was used in training because the accuracy on it would be substantially better. Moreover, since JB is the \emph{only} speaker for whom the models have seen spontaneous speech, these models will be unusually accurate on JB and substantially less accurate on other speakers. An astute observer would then have no trouble concluding that JB was used as training data. In other words, the observer will have performed a membership inference attack. 

In the third row of the table, we have used a model that had a warm start with public data, and then was trained through federated learning and the various differential privacy techniques described earlier. Accuracy on the Librispeech dataset has improved compared to the baseline on the first row, albeit not as significantly as using the data openly. The real benefit though is the performance on the JB dataset. The accuracy on the JB dataset is about the same as the first row where JB is not used in training at all. \emph{In regards to privacy, this is exactly what we wanted.} Recall that JB was designed to be a prominent outlier, and to look very different from the other datasets. Our DP-based technique is designed to strongly prefer characteristics in the data that are common to everyone, and discard characteristics that belong to few. Even when trained with very large privacy parameters, the JB dataset remained indistinctive, as indicated by the low accuracy scores on the third row.

\section{Discussion}
While we successfully demonstrated the possibility of building a privacy-preserving acoustic model using DP, there are complications with hyper-parameter optimization.
In our case, we rely heavily on the public dataset for hyperparameter tuning.
It is possible to exploit the public data to infer hyperparameters that might perform well in private training.
An approach is to hold out a split of the public data, and then simulate mock private training to find feasible clipping bounds, step sizes, ideal effective batch sizes and privacy budget allocations for private training.
Another approach to hyper-parameter tuning is to train many private models with different combinations of hyper-parameters, and then invoke the exponential mechanism to approximately choose and release the model with the lowest loss or highest accuracy.
Unfortunately, the privacy loss involved with this approach scales linearly with the number of models tried; you must account for DP training of all models, not just the approximately best model. 
Liu at al. provide algorithms that reduce this to a constant loss in privacy\cite{Liu19Psel}.
Hyper-parameter tuning is a primary topic of interest for future work.


Differential Privacy is a powerful idea and DP-based techniques have been used for a wide variety of applications. DP techniques provide stronger privacy protections and more accurate results when the number of individuals in the dataset gets larger. It is easy to hide in a crowd; less so in a group of three. Unlike a lot of different tasks, speech has a problem here. Collecting speech data from a large number of unique individuals is particularly difficult. Datasets tend to contain many hours of speech from a small number of individuals, not short clips of speech from millions of different individuals. 
Contrast this to a problem like text processing where collecting text written by a million different people is relatively straightforward. 
The limited availability of such data hinders the use of differential privacy in this space. 
On the other hand, this is the kind of difficulty that federated learning is best-suited to address. 

\section{Acknowledgements}

This material is based upon work supported by the Defense Advanced Research Projects Agency (DARPA) under Agreement No. HR00112090105.  Any opinions, findings and conclusions or recommendations expressed in this material are those of the author(s) and do not necessarily reflect the views of the United States Government or DARPA.

\bibliographystyle{IEEEtran}

\bibliography{mybib}

\begin{thebibliography}{10}
\providecommand{\url}[1]{#1}
\csname url@samestyle\endcsname
\providecommand{\newblock}{\relax}
\providecommand{\bibinfo}[2]{#2}
\providecommand{\BIBentrySTDinterwordspacing}{\spaceskip=0pt\relax}
\providecommand{\BIBentryALTinterwordstretchfactor}{4}
\providecommand{\BIBentryALTinterwordspacing}{\spaceskip=\fontdimen2\font plus
\BIBentryALTinterwordstretchfactor\fontdimen3\font minus
  \fontdimen4\font\relax}
\providecommand{\BIBforeignlanguage}[2]{{%
\expandafter\ifx\csname l@#1\endcsname\relax
\typeout{** WARNING: IEEEtran.bst: No hyphenation pattern has been}%
\typeout{** loaded for the language `#1'. Using the pattern for}%
\typeout{** the default language instead.}%
\else
\language=\csname l@#1\endcsname
\fi
#2}}
\providecommand{\BIBdecl}{\relax}
\BIBdecl

\bibitem{kairouz2021advances}
P.~Kairouz, H.~B. McMahan, B.~Avent, A.~Bellet, M.~Bennis, A.~N. Bhagoji,
  K.~Bonawitz, Z.~Charles, G.~Cormode, R.~Cummings \emph{et~al.}, ``Advances
  and open problems in federated learning,'' \emph{Foundations and
  Trends{\textregistered} in Machine Learning}, vol.~14, no. 1--2, pp. 1--210,
  2021.

\bibitem{black2015random}
A.~W. Black and P.~K. Muthukumar, ``Random forests for statistical speech
  synthesis,'' in \emph{Sixteenth Annual Conference of the International Speech
  Communication Association}, 2015.

\bibitem{dwork2014algorithmic}
C.~Dwork, A.~Roth \emph{et~al.}, ``The algorithmic foundations of differential
  privacy.'' \emph{Found. Trends Theor. Comput. Sci.}, vol.~9, no. 3-4, pp.
  211--407, 2014.

\bibitem{smaragdis2007framework}
P.~Smaragdis and M.~Shashanka, ``A framework for secure speech recognition,''
  \emph{IEEE Transactions on Audio, Speech, and Language Processing}, vol.~15,
  no.~4, pp. 1404--1413, 2007.

\bibitem{pathak2011privacy}
M.~Pathak, S.~Rane, W.~Sun, and B.~Raj, ``Privacy preserving probabilistic
  inference with hidden markov models,'' in \emph{2011 IEEE International
  Conference on Acoustics, Speech and Signal Processing (ICASSP)}.\hskip 1em
  plus 0.5em minus 0.4em\relax IEEE, 2011, pp. 5868--5871.

\bibitem{pathak2011privacya}
M.~A. Pathak and B.~Raj, ``Privacy preserving speaker verification using
  adapted gmms,'' in \emph{Twelfth Annual Conference of the International
  Speech Communication Association}, 2011.

\bibitem{pathak2013privacy}
M.~A. Pathak, B.~Raj, S.~D. Rane, and P.~Smaragdis, ``Privacy-preserving speech
  processing: cryptographic and string-matching frameworks show promise,''
  \emph{IEEE signal processing magazine}, vol.~30, no.~2, pp. 62--74, 2013.

\bibitem{ahmed2020preech}
S.~Ahmed, A.~R. Chowdhury, K.~Fawaz, and P.~Ramanathan, ``Preech: a system for
  privacy-preserving speech transcription,'' in \emph{Proceedings of the 29th
  USENIX Conference on Security Symposium}, 2020, pp. 2703--2720.

\bibitem{tomashenko2020introducing}
N.~Tomashenko, B.~M.~L. Srivastava, X.~Wang, E.~Vincent, A.~Nautsch,
  J.~Yamagishi, N.~Evans, J.~Patino, J.-F. Bonastre, P.-G. No{\'e}
  \emph{et~al.}, ``Introducing the voiceprivacy initiative,'' in
  \emph{INTERSPEECH}, 2020.

\bibitem{tomashenko2022voiceprivacy}
N.~Tomashenko, X.~Wang, E.~Vincent, J.~Patino, B.~M.~L. Srivastava, P.-G.
  No{\'e}, A.~Nautsch, N.~Evans, J.~Yamagishi, B.~O’Brien \emph{et~al.},
  ``The voiceprivacy 2020 challenge: Results and findings,'' \emph{Computer
  Speech \& Language}, vol.~74, p. 101362, 2022.

\bibitem{wood2018differential}
A.~Wood, M.~Altman, A.~Bembenek, M.~Bun, M.~Gaboardi, J.~Honaker, K.~Nissim,
  D.~R. O'Brien, T.~Steinke, and S.~Vadhan, ``Differential privacy: A primer
  for a non-technical audience,'' \emph{Vand. J. Ent. \& Tech. L.}, vol.~21, p.
  209, 2018.

\bibitem{gaboardi2020programming}
M.~Gaboardi, M.~Hay, and S.~Vadhan, ``A programming framework for open{DP},''
  \emph{Manuscript}, 2020.

\bibitem{abadi2016deep}
M.~Abadi, A.~Chu, I.~Goodfellow, H.~B. McMahan, I.~Mironov, K.~Talwar, and
  L.~Zhang, ``Deep learning with differential privacy,'' in \emph{Proceedings
  of the 2016 ACM SIGSAC conference on computer and communications security},
  2016, pp. 308--318.

\bibitem{paszke2019pytorch}
A.~Paszke, S.~Gross, F.~Massa, A.~Lerer, J.~Bradbury, G.~Chanan, T.~Killeen,
  Z.~Lin, N.~Gimelshein, L.~Antiga \emph{et~al.}, ``Pytorch: An imperative
  style, high-performance deep learning library,'' \emph{Advances in neural
  information processing systems}, vol.~32, 2019.

\bibitem{opendpurl}
\url{https://github.com/opendp/opendp/}.

\bibitem{hwang1992subphonetic}
M.-Y. Hwang and X.~Huang, ``Subphonetic modeling with markov states-senone,''
  in \emph{ICASSP-92: 1992 IEEE International Conference on Acoustics, Speech,
  and Signal Processing}, vol.~1.\hskip 1em plus 0.5em minus 0.4em\relax IEEE,
  1992, pp. 33--36.

\bibitem{panayotov2015librispeech}
V.~Panayotov, G.~Chen, D.~Povey, and S.~Khudanpur, ``Librispeech: an asr corpus
  based on public domain audio books,'' in \emph{2015 IEEE international
  conference on acoustics, speech and signal processing (ICASSP)}.\hskip 1em
  plus 0.5em minus 0.4em\relax IEEE, 2015, pp. 5206--5210.

\bibitem{Liu19Psel}
\BIBentryALTinterwordspacing
J.~Liu and K.~Talwar, ``Private selection from private candidates,'' in
  \emph{Proceedings of the 51st Annual ACM SIGACT Symposium on Theory of
  Computing}, ser. STOC 2019.\hskip 1em plus 0.5em minus 0.4em\relax New York,
  NY, USA: Association for Computing Machinery, 2019, p. 298–309. [Online].
  Available: \url{https://doi.org/10.1145/3313276.3316377}
\BIBentrySTDinterwordspacing

\end{thebibliography}


\end{document}